\documentclass[11pt]{article}
\usepackage{colortbl}

\usepackage[final]{acl}

\usepackage{times}
\usepackage{latexsym}

\usepackage[T1]{fontenc}

\usepackage[utf8]{inputenc}

\usepackage{microtype}

\usepackage{inconsolata}

\usepackage{graphicx}
\usepackage{subcaption}

\usepackage{amsmath,amssymb,amsfonts}
\usepackage{algorithmic}
\usepackage{graphicx}
\usepackage{textcomp}
\usepackage{xcolor}
\usepackage{url}
\usepackage{xurl}

\usepackage{listings}
\lstdefinelanguage{JavaScript}{
  keywords={break, case, catch, continue, debugger, default, delete, do, else, finally, for, function, if, in, instanceof, new, return, switch, this, throw, try, typeof, var, void, while, with, let, const},
  keywordstyle=\color{blue}\bfseries,
  ndkeywords={class, export, boolean, throw, implements, import, this},
  ndkeywordstyle=\color{darkgray}\bfseries,
  identifierstyle=\color{black},
  sensitive=false,
  comment=[l]{//},
  morecomment=[s]{/*}{*/},
  commentstyle=\color{gray}\ttfamily,
  stringstyle=\color{red}\ttfamily,
  morestring=[b]',
  morestring=[b]"
}
\usepackage{color}
\usepackage{xspace}
\usepackage{algorithm}
\usepackage{booktabs}
\usepackage{xltabular}
\usepackage{wrapfig}
\usepackage{float}      
\usepackage{caption}    

\usepackage{hyperref}   
\usepackage{hyperxmp}
\usepackage{microtype}
\usepackage{multirow}

\usepackage{tcolorbox}

\newcommand{\TheName}{\textsc{EyeMulator}}

\def\BibTeX{{\rm B\kern-.05em{\sc i\kern-.025em b}\kern-.08em
    T\kern-.1667em\lower.7ex\hbox{E}\kern-.125emX}}

\begin{document}


\title{\TheName{}: Improving Code Language Models by Mimicking \\Human Visual Attention}


\author{
    \textbf{Yifan Zhang}\textsuperscript{1}\hspace{5pt}
    \textbf{Chen Huang}\textsuperscript{2}\hspace{5pt}
    \textbf{Yueke Zhang}\textsuperscript{1}\hspace{5pt}
    \textbf{Jiahao Zhang}\textsuperscript{1}\\
    \textbf{Toby Jia-Jun Li}\textsuperscript{3}\hspace{5pt}
    \textbf{Collin McMillan}\textsuperscript{3}\hspace{5pt}
    \textbf{Kevin Leach}\textsuperscript{1}\hspace{5pt}
    \textbf{Yu Huang}\textsuperscript{1}\vspace{5pt}\\
    Vanderbilt University\textsuperscript{1}\hspace{5pt}
    National University of Singapore\textsuperscript{2}\hspace{5pt}
    University of Notre Dame\textsuperscript{3}\\
    \normalsize
    \texttt{\{yifan.zhang.2,yueke.zhang,jiahao.zhang,kevin.leach,yu.huang\}@vanderbilt.edu}\\
    \normalsize
    \texttt{huang\_chen@nus.edu.sg}\hspace{5pt}
    \normalsize
    \texttt{\{toby.j.li,cmc\}@nd.edu}
}

\maketitle

\begin{abstract}
Code Language Models (CodeLLMs) learn token importance from data correlations, whereas human developers attend selectively to semantically salient code. We present \TheName{}, a model-agnostic method that injects human visual-attention priors into CodeLLM fine-tuning without architectural changes. \TheName{} distills eye-tracking data into semantic salience and gaze-transition priors, then uses them to reweight token-level training losses. Across six backbones, two data regimes, and three CodeXGLUE tasks, the reported configurations yield positive matched-metric deltas in all 36 model-task-setting cells. Effects are largest for structure-preserving completion and translation, while summarization shows smaller but positive METEOR deltas. Session-mode and component-ablation analyses further show that reading, writing, semantic, and transition-derived priors provide complementary signal. Human-attention artifacts are available at \url{https://zenodo.org/records/17205682}.
\end{abstract}

\begin{figure*}[t]
  \centering
  \includegraphics[width=\textwidth]{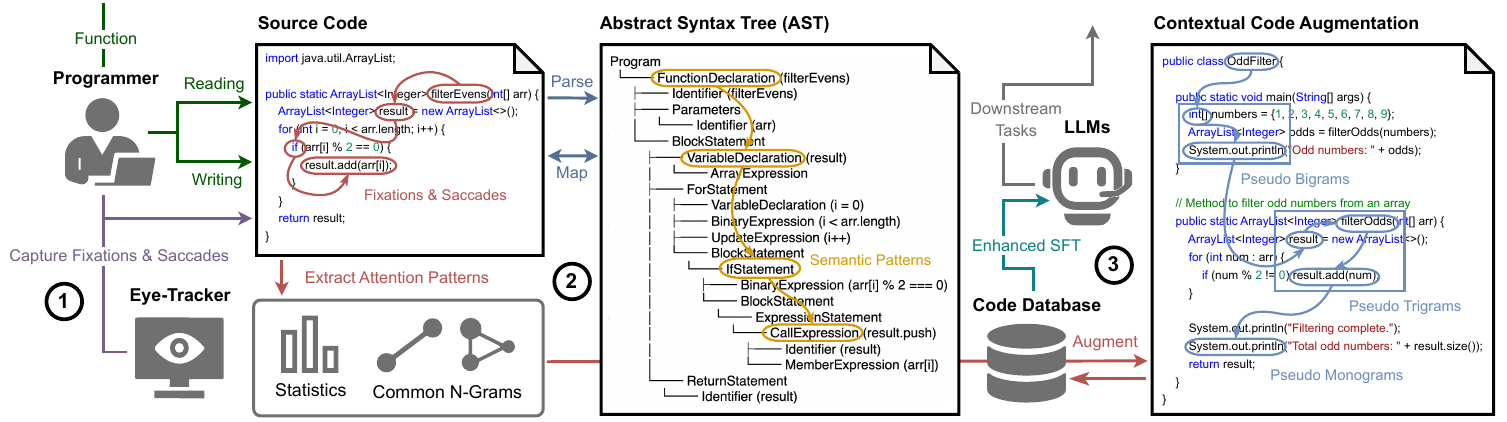}
  \caption{Overview of \TheName{}. Human gaze traces are aligned to AST tokens, distilled into semantic salience and transition priors, and converted into pseudo-scan paths that guide weighted SFT and token-level preference optimization without changing the base CodeLLM architecture.}
  \label{fig:attention_pipeline}
\end{figure*}

\section{Introduction}
\label{sec:introduction}

Large Language Models (LLMs) have fundamentally altered the landscape of software engineering, demonstrating exceptional capability in tasks ranging from automated code generation to bug localization. These models, typically based on the Transformer architecture, learn to predict tokens by minimizing loss over internet-scale repositories~\cite{vaswani2017attention}. However, this process relies on ``machine attention,'' a statistical mechanism that treats context uniformly based on data correlations. In contrast, human developers employ distinct cognitive strategies characterized by program comprehension~\cite{obrien2003software,harth2017program}. Eye-tracking research consistently demonstrates that experts rely on intuition to form mental models. They fixate selectively on semantically critical elements, such as control flow conditions and method signatures, while skimming over syntactic sugar and boilerplate~\cite{sharafi_practical_2020,huang2020biases}.


The dominant paradigm for adapting CodeLLMs relies heavily on minimizing prediction error over vast datasets or aligning models via high-level instruction tuning. While these methods improve general adherence to user intent, they do not fundamentally alter the model's underlying attention mechanism, which remains tethered to statistical co-occurrences rather than semantic reasoning. Existing attempts to bridge this gap, such as retrieval-augmented generation (RAG), provide external context but fail to teach the model \textit{how} to process that context like an expert. Consequently, even state-of-the-art models continue to distribute attention uniformly across boilerplate and logic, missing the nuanced, selective focus that characterizes human program comprehension.

To bridge the critical gap between statistical machine attention and selective human focus, we propose \TheName{}, a methodology to ground CodeLLM training in cognitive visual patterns. We posit that while purely data-driven models excel at surface-level pattern matching, they fundamentally lack the ability to prioritize the logical dependencies, such as variable data flow and execution paths, that characterize expert comprehension. By explicitly aligning model attention with human scan paths, we enable the system to process code with a semantic salience mirroring that of a developer. Crucially, unlike prior gaze-aware approaches that necessitate specialized architectures or expensive pre-training from scratch~\cite{zhang2024eyetrans}, we aim to distill these cognitive insights into a modular, model-agnostic signal that can be seamlessly integrated into standard supervised fine-tuning pipelines.

The core innovation of \TheName{} is a generative attention mechanism that allows standard CodeLLMs to learn human cognitive patterns directly from text. We achieve this by first distilling raw eye-tracking data into two portable artifacts: \textit{Semantic Salience Priors}, which model the static importance of token types (e.g., variables vs. keywords), and \textit{Transition Probabilities}, which capture the dynamic flow of expert reading (e.g., jumping from definition to usage). These artifacts enable us to synthesize ``pseudo-scan paths'' for standard training data where no eye-tracking exists. We then align the model to these paths using a composite objective: a re-weighted Supervised Fine-Tuning (SFT) loss to emphasize salient tokens, and a token-level Direct Preference Optimization (DPO) loss that explicitly rewards the model for prioritizing human-preferred context over irrelevant boilerplate.

We evaluate \TheName{} on three diverse CodeXGlue tasks: code completion, Java-to-C\# translation, and Java-to-English summarization. To test robustness, we run six small CodeLLM backbones across low-data and full-data regimes, with seed 42 and seed 12345 used for robustness checks. The cleaned results show positive quality-safe gains in every model--task--setting cell. Completion and translation provide the strongest evidence, with consistent Exact Match improvements across all six models and both data regimes; summarization is more recipe-sensitive but remains positive under METEOR after targeted tuning and DeepSeek tokenizer correction. Session-mode, component-ablation, and attention diagnostics further support the mechanism behind these gains.


In summary, this paper contributes: (1) a portable pipeline for converting raw human gaze traces into reusable semantic salience and sequential transition priors over code tokens; (2) a model-agnostic fine-tuning objective that incorporates these priors through token-level weighting and preference alignment without requiring architectural changes; and (3) a six-backbone evaluation across multiple CodeXGlue tasks and data regimes, showing that gaze-derived signals provide consistent quality-safe gains while revealing when human-attention cues are robust, task-sensitive, and complementary.

\section{Approach}
\label{sec:approach}

As illustrated in Figure~\ref{fig:attention_pipeline}, \TheName{} consists of three stages: processing raw gaze data, extracting statistical attention priors, and fine-tuning the model using a dual-objective loss.

\subsection{Data Transformation and Metrics}
\begin{figure}[t]
    \centering
    \includegraphics[width=\columnwidth]{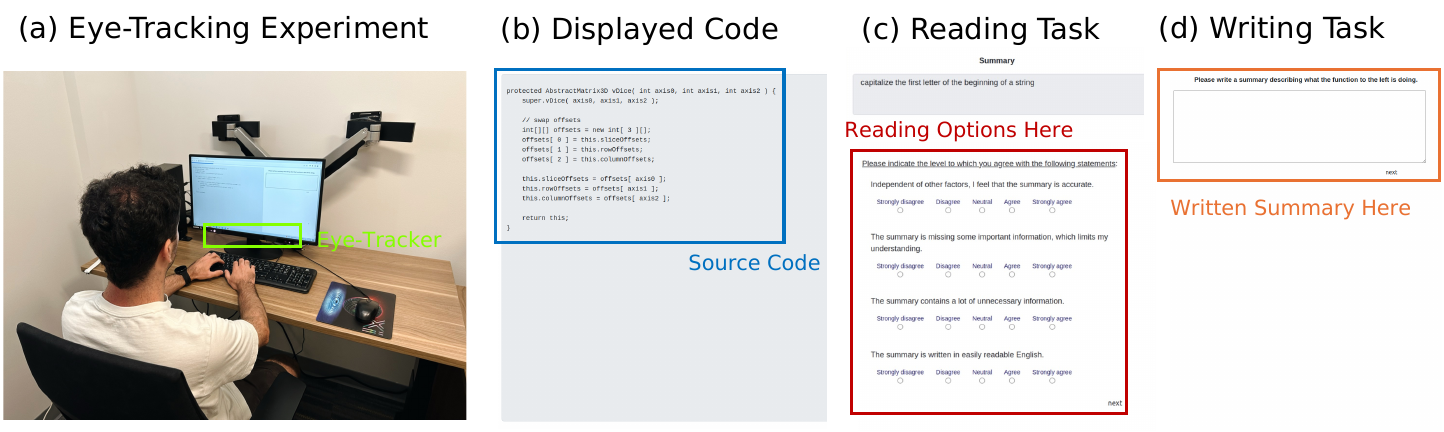}
    \caption{EyeTrans gaze data used to derive human-attention priors. Fixations over Java snippets are aligned with AST tokens and separated into reading and writing sessions before statistical distillation.}
    \label{fig:study_design}
\end{figure}

We leverage the EyeTrans corpus~\cite{zhang2024eyetrans}, which contains 120Hz gaze recordings from 27 programmers reading and writing Java code. To transform raw sensor data into a training signal, we first apply a dispersion-threshold (I-DT) algorithm to identify \textbf{Fixations} ($\mathcal{F}$), defined as stable gaze points lasting at least 100 ms. These fixations serve as our primary metric of cognitive interest. We then spatially map these fixations to Abstract Syntax Tree (AST) leaf tokens using bounding boxes captured during the study (Figure~\ref{fig:study_design}). Unmapped points, which account for approximately 3\% of the data, are discarded. This process yields 1,565 \textbf{Scan Paths} ($\mathcal{P}$), representing the chronological sequence of attended tokens aligned with the code structure.

\subsection{Attention Pattern Extraction}
\label{subsec:pattern_extraction}
We distill the token-aligned fixations into two statistical artifacts: semantic salience priors and sequential transition models.

\paragraph{Semantic Salience Priors.}
To model the inherent importance of different token types (e.g., \texttt{ForStatement} vs. \texttt{Identifier}), we employ a Bayesian approach. For each semantic class $s$, we count the number of fixations $c_1(s)$ relative to the total number of available tokens $n_s^{\mathrm{tok}}$. We model the probability of attention $\theta_s$ using a Beta distribution $\mathrm{Beta}(\alpha_s, \beta_s)$, which serves as a conjugate prior to the binomial likelihood of fixation. We set $\alpha_s = c_1(s) + 1$ and $\beta_s = n_s^{\mathrm{tok}} - c_1(s) + 1$. The posterior mean $\mathbb{E}[\theta_s] = \alpha_s / (\alpha_s + \beta_s)$ provides a robust estimate of salience, smoothing out noise in rare token classes.

\paragraph{Sequential Transition Models.}
To capture the flow of expert reading, we count bigrams and trigrams within the scan paths. We compute conditional probabilities $P(s_b|s_a)$ and $P(s_c|s_a,s_b)$, which represent the likelihood of transitioning between specific semantic states (e.g., from a variable definition to its usage). We prune n-grams with fewer than 5 occurrences to reduce noise.

\subsection{Pseudo-Attention Path Generation}
\begin{figure}[t]
  \centering
  \includegraphics[width=\linewidth]{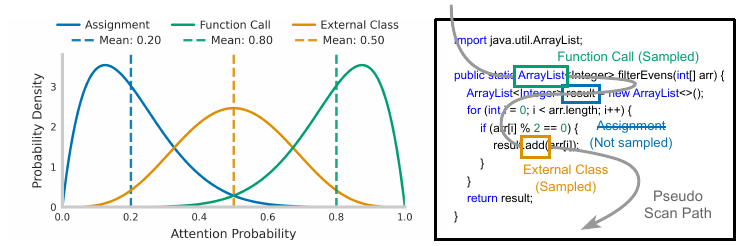}
  \caption{Pseudo-scan-path generation. Semantic salience priors identify important code tokens, while transition statistics connect them into gaze-like token paths for training supervision.}
  \label{fig:eyemulator_pseudo_path}
\end{figure}

To simulate human attention for standard training data where no eye-tracking exists, we generate ``pseudo scan paths'' $\tilde{\mathcal{P}}$. For an input sequence of length $n$, we proceed in three steps: (1) We sample a saliency ratio $\rho$ from the aggregate Beta distribution of the corpus. (2) We determine the total count of attended tokens $m = \lfloor \rho n \rfloor$ and allocate quotas to specific token classes based on their posterior means $\mathbb{E}[\theta_s]$. (3) We construct the path $\tilde{\mathcal{P}}$ by greedily matching masked tokens into valid n-grams (prioritizing Trigrams, then Bigrams, then Monograms) that satisfy a line-span constraint $L$. This procedure (Figure~\ref{fig:eyemulator_pseudo_path}) synthesizes sequences that preserve both the local structure and the semantic selectivity of human attention.

\subsection{Gaze-Informed Fine-Tuning}
\label{subsec:gaze_informed_finetuning}
We seamlessly integrate these artifacts into standard LLM training using a novel weighting scheme and a composite loss function.

\paragraph{Weight Calculation.}
Since LLMs typically use subword tokenizers (e.g., Byte-Pair Encoding), we project our AST-level weights by assigning the parent token's weight to all its constituent subword shards. We compute a final training weight $w_j$ for each token $j$ by combining a base importance term, an inverse frequency term to upweight rare code constructs, and the semantic probability: $w_j = w_{\text{base}} + (\log(\text{freq}(g_j)+2))^{-1} + \mathbb{E}[\theta_{s_j}]$.

\paragraph{Composite Objective.}
We fine-tune the model using a loss function $\mathcal{L}(\phi) = \mathcal{L}_{\mathrm{SFT}}(\phi) + \gamma \mathcal{L}_{\mathrm{DPO}}(\phi)$.
The \textbf{Weighted SFT Loss} is a modification of the standard categorical cross-entropy, scaled by the calculated weights: $\mathcal{L}_{\mathrm{SFT}}(\phi) = - \sum_{j} w_j \log P_\phi(x_j | x_{<j})$.
The \textbf{Token-Level DPO Loss} adapts the Direct Preference Optimization framework~\cite{rafailovDirectPreferenceOptimization2024a} to our setting. DPO typically optimizes a policy to prefer a winning sample $y_w$ over a losing sample $y_l$. We define our ``winning'' trajectory as the generated pseudo-scan path (high salience) and the ``losing'' trajectory as the complement (low salience background tokens). This term explicitly rewards the model for assigning higher probability to the semantically salient tokens that humans prioritize.

\subsection{Integrated Training Procedure}
\label{subsec:integrated_algorithm}

We synthesize the artifact extraction, path generation, and loss computation into a unified training loop. Algorithm~\ref{alg:eyemulator_training} details the complete fine-tuning procedure.

The process begins by initializing the model $\phi$ and loading the distilled EyeTrans artifacts (Semantic Priors $\mathbb{E}[\theta]$ and Transition Probabilities $P_{\mathrm{trans}}$). For every batch of code sequences, we dynamically generate pseudo-scan paths $\tilde{\mathcal{P}}$ that mimic human visual attention. These paths serve as the ground truth for calculating token-specific importance weights $W$. Finally, the model parameters are updated via gradient descent on the composite objective $\mathcal{L}_{\mathrm{total}}$, which balances the reconstruction of salient tokens ($\mathcal{L}_{\mathrm{SFT}}$) with the preference alignment of attention trajectories ($\mathcal{L}_{\mathrm{DPO}}$).

\begin{algorithm}[t]
\caption{\TheName{} Integrated Fine-Tuning}
\label{alg:eyemulator_training}
\begin{small}
\textbf{Input:} Pre-trained CodeLLM $\phi_0$, Code Dataset $\mathcal{D} = \{x^{(i)}\}_{i=1}^{N}$, EyeTrans Artifacts (Semantic Priors $\mathbb{E}[\theta]$, Transitions $P_{\mathrm{trans}}$) \\
\textbf{Hyperparameters:} DPO weight $\gamma$, Learning rate $\eta$ \\
\textbf{Output:} Fine-tuned Model $\phi^*$

\begin{algorithmic}[1] 
\STATE $\phi \leftarrow \phi_0$
\WHILE{not converged}
    \STATE Sample batch $\mathcal{B} \sim \mathcal{D}$
    \FORALL{sequence $x \in \mathcal{B}$}
        \STATE \textit{// Stage 1: Pseudo-Path Generation (Sec.~\ref{subsec:pattern_extraction})}
        \STATE $Tags \leftarrow \text{GetASTTags}(x)$
        \STATE $\rho \sim \text{Beta}(\alpha_{\text{agg}}, \beta_{\text{agg}})$ \COMMENT{Sample attention density}
        \STATE $\tilde{\mathcal{P}} \leftarrow \text{GeneratePath}(x, Tags, \rho, \mathbb{E}[\theta], P_{\mathrm{trans}})$
        
        \STATE \textit{// Stage 2: Weight Calculation (Sec.~\ref{subsec:gaze_informed_finetuning})}
        \FORALL{token $x_j \in x$}
            \STATE $w_j \leftarrow w_{\text{base}} + \frac{1}{\log(\text{freq}(x_j)+2)} + \mathbb{E}[\theta_{\text{tag}(x_j)}]$
        \ENDFOR
        \STATE $W \leftarrow \{w_j\}_{j=1}^{|x|}$
        
        \STATE \textit{// Stage 3: Dual-Objective Optimization}
        \STATE $\mathcal{L}_{\mathrm{SFT}} \leftarrow - \sum_{j \in \tilde{\mathcal{P}}} w_j \log P_\phi(x_j | x_{<j})$
        \STATE $\mathcal{L}_{\mathrm{DPO}} \leftarrow \text{DPOLoss}(\phi, \tilde{\mathcal{P}}, x \setminus \tilde{\mathcal{P}}, W)$
        \STATE $\mathcal{L}_{\mathrm{total}} \leftarrow \mathcal{L}_{\mathrm{SFT}} + \gamma \mathcal{L}_{\mathrm{DPO}}$
    \ENDFOR
    \STATE $\phi \leftarrow \phi - \eta \nabla_\phi \sum_{x \in \mathcal{B}} \mathcal{L}_{\mathrm{total}}$
\ENDWHILE
\RETURN $\phi$
\end{algorithmic}
\end{small}
\end{algorithm}

\section{Experimental Setup}
\label{sec:experimental_setup}

We design our experiments to evaluate the efficacy of gaze-informed training across different code intelligence tasks and model architectures. We specifically address five research questions: (RQ1) How faithfully do our distilled artifacts capture human attention patterns? (RQ2) Does mimicking human attention improve performance on downstream code-intelligence tasks? (RQ3) How do reading-derived versus writing-derived priors differ in their impact across tasks? (RQ4) How do the individual components of \TheName{} contribute to overall gains? (RQ5) Does the fine-tuned model actually learn to attend to semantically salient regions?

\paragraph{Datasets and Benchmarks.}
We utilize the EyeTrans corpus~\cite{zhang2024eyetrans} solely for extracting attention artifacts as detailed in Section~\ref{sec:approach}. For downstream evaluation, we employ CodeXGlue~\cite{lu2021codexglue}, selecting three tasks to test generalization. For \textit{Code Translation} (Java to C\#), we use the standard Java-to-C\# split. For \textit{Code Summarization} (Java to English), we use 16,490 training, 518 validation, and 1,095 test examples. For \textit{Code Completion}, we sample 20\% of the CodeXGlue Java completion corpus, yielding 1,633 training, 1,005 validation, and 1,318 test examples. We report two data regimes: a low-data setting with 200 training examples and a full-data setting using the available training split. Completion and translation are evaluated with Exact Match, while summarization is evaluated with METEOR. Appendix~\ref{sec:appendix_metrics} provides the metric definitions.

\paragraph{Models and Baselines.}
To demonstrate model agnosticism, we apply \TheName{} to six backbones: StarCoderBase-1B, Llama-3.2-1B, DeepSeek-Coder-1.3B, Qwen2.5-Coder-1.5B, SmolLM3-3B, and Granite-Code-2B. Each baseline is fine-tuned on the same task data without gaze weights. \TheName{} uses task-level LoRA/QLoRA recipes, with code completion and translation using a rank-32, weight-4 recipe and summarization using lower-rank recipes selected from completed quality-safe variants. For robustness, we run the clean recipe on seeds 42 and 12345 and report best quality-safe results only when generated outputs pass qualitative sanity checks.

\paragraph{Implementation Details.} 
All experiments use HuggingFace \texttt{transformers} with parameter-efficient LoRA/QLoRA adapters. Input sequences are processed with a maximum length of 1024 tokens. Completion and translation use rank-32 LoRA with gaze weight 4 as the clean recipe; summarization uses lower-rank recipes selected from quality-safe sweeps. We use AdamW-style optimization, task-specific learning rates, and seeds 42 and 12345 for robustness checks. Detailed training configurations and attention-prior alignment steps are documented in Appendix~\ref{sec:appendix_implementation}.

\section{Result Analysis}
\label{sec:result_analysis}

We organize the results from broadest to most diagnostic. We first establish the main effect across backbones, tasks, and data regimes (RQ2), then use session-mode and component-ablation analyses to explain which gaze signals matter (RQ3--RQ4). Finally, we retain the attention-map analysis as an appendix diagnostic rather than as the primary evidence (RQ5).

\subsection{RQ1: Artifact Distillation}
\label{subsec:rq1}

We assess how well distilled artifacts capture human attention by examining n-gram fixation patterns, fitted Beta parameters for semantic labels, and the resulting attention distributions across reading, writing, and combined tasks.

\paragraph{N-gram Analysis.}
To assess the consistency and structure of human attention, we mapped programmers' fixation sequences to semantic categories and extracted the most frequent transitions. As shown in Table~\ref{tab:ngram_counts}, the resulting patterns reveal a non-linear yet highly organized reading strategy. The high count of \textit{function declaration $\to$ variable declaration} bigrams (8,399) indicates that developers frequently switch focus between interface definitions and their implementation details. Similarly, the \textit{conditional statement $\to$ loop} pattern (6,026) reflects an iterative inspection of branching logic and control flow. These recurring transition motifs demonstrate that expert attention is driven by logical dependencies rather than linear scanning.

\begin{table}[t]
\centering
\caption{Representative gaze patterns extracted from token-aligned fixations. Frequent transitions show that programmers repeatedly move between declarations, parameters, and control-flow tokens rather than scanning code uniformly.}
\resizebox{\columnwidth}{!}{%
\begin{tabular}{llr}
\toprule
\textbf{Type} & \textbf{Sequence} & \textbf{Count} \\
\midrule
Mono & variable declaration & 18665 \\
Mono & conditional statement & 13222 \\
\cmidrule{1-3}
Bigram & function decl $\to$ variable decl & 8399 \\
Bigram & conditional statement $\to$ loop & 6026 \\
\cmidrule{1-3}
Trigram & func decl $\to$ param $\to$ var decl & 1634 \\
Trigram & conditional $\to$ func decl $\to$ param & 1199 \\
\bottomrule
\end{tabular}}
\label{tab:ngram_counts}
\end{table}

\begin{figure}[ht]
    \centering
    \includegraphics[width=\columnwidth]{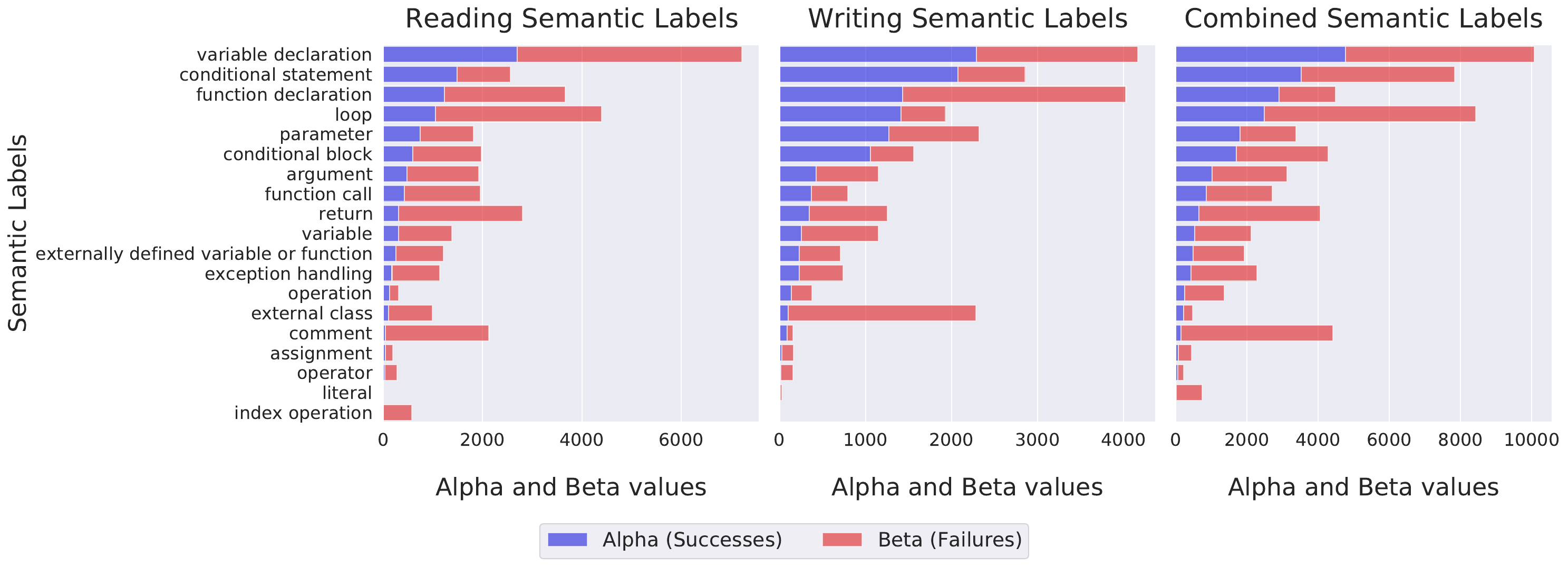}
    \caption{Estimated Beta parameters for semantic token classes. Larger $\alpha_s$ values indicate token types that consistently attract developer fixations, with variable declarations emerging as a primary attention target.}
    \label{fig:beta_distributions}
\end{figure}

\paragraph{Beta Parameter Estimation.}
Figure~\ref{fig:beta_distributions} presents the fitted Beta parameters ($\alpha_s=$ gaze hits, $\beta_s=$ gaze misses) for each semantic label. Consistent with the n-gram findings, \texttt{variable declaration} exhibits the highest $\alpha_{s}$ in both reading and combined tasks, reinforcing its role as a primary cognitive anchor for developers. In contrast, control-flow labels like \texttt{loop} show more balanced ratios ($\alpha_{s}=7,876$, $\beta_{s}=4,876$), indicating that attention to these constructs is more context-dependent, shifting between cursory scanning and deeper inspection depending on the complexity of the logic.

\begin{figure}[ht]
  \centering
  \includegraphics[width=\columnwidth]{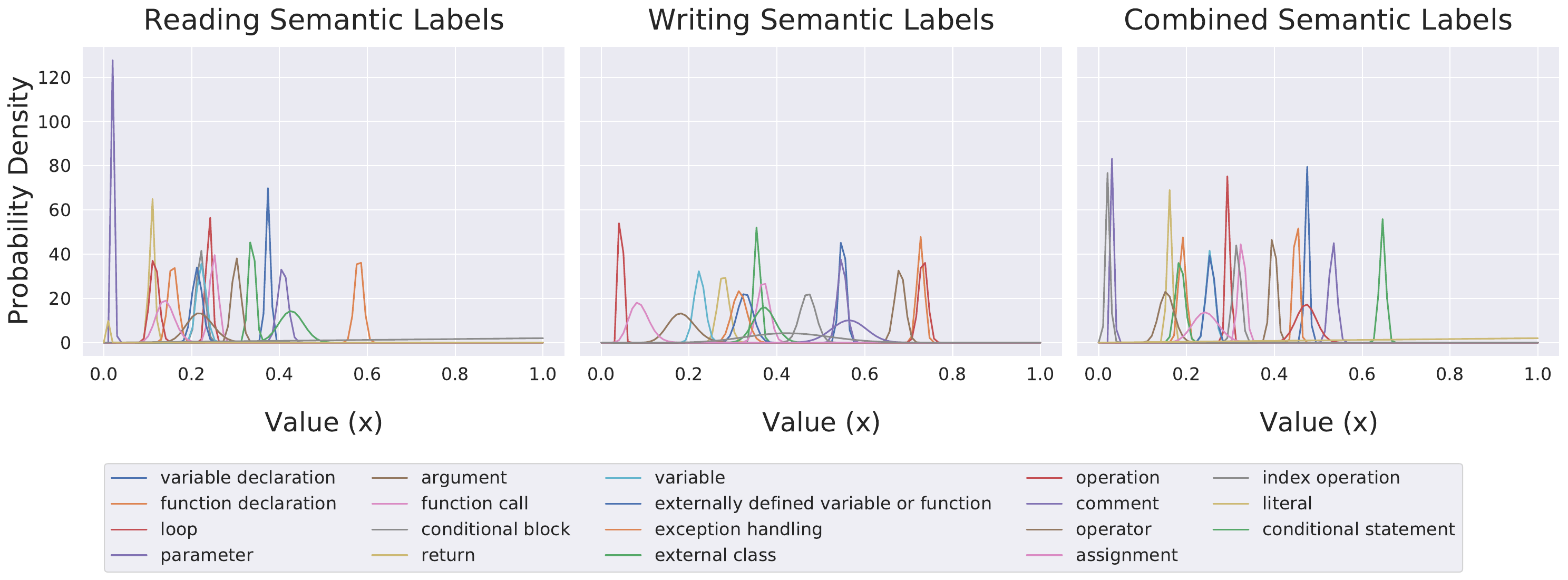}
  \caption{Smoothed Beta density functions for representative semantic classes. Narrow peaks indicate consistently focused attention, while broader curves capture context-dependent inspection of control-flow constructs.}
  \label{fig:beta_curves}
\end{figure}

\paragraph{Density Function Visualization.}
Figure~\ref{fig:beta_curves} illustrates the Beta distribution density functions derived from these parameters. Reading-only curves are sharply peaked; for instance, \texttt{variable declaration} centers tightly around a high probability, signifying focused and reliable inspection. Conversely, the distributions for \texttt{conditional statement} in the combined task are broader and even bimodal. This variance suggests divergent reading strategies, where developers may either skim standard conditions quickly or fixate heavily on complex predicates, confirming that our artifacts capture the nuance of cognitive load.

\subsection{RQ2: Main Cross-Task Results}
\label{subsec:rq2}

\begin{figure}[t]
  \centering
  \includegraphics[width=\columnwidth]{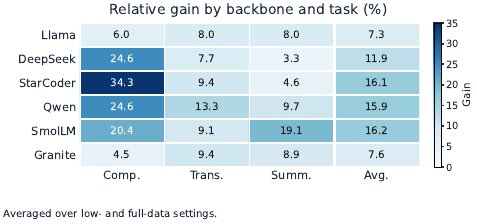}
  \caption{Backbone-by-task profile of \TheName{} gains. Each cell averages relative improvement over low- and full-data settings; darker colors and annotations indicate larger improvements.}
  \label{fig:rq2_model_task_profile}
\end{figure}

RQ2 asks whether gaze-derived priors improve downstream CodeLLM fine-tuning across tasks, data regimes, and backbones. The main result is that \TheName{} improves every evaluated cell under the matched task metric: 36 positive gains across six backbones, three tasks, and two data settings. Table~\ref{tab:rq2_main_improvements} reports the selected quality-safe run for each cell, and Figure~\ref{fig:rq2_model_task_profile} summarizes the same evidence as a model-by-task profile.

\begin{table*}[t]
\raggedright
\captionsetup{justification=raggedright,singlelinecheck=false}
\scriptsize
\caption{Main cross-task evaluation across six backbones, two data regimes, and three CodeXGlue tasks. Gray rows report \TheName{} scores; parentheses show relative improvement over the matched baseline under the task metric.}
\label{tab:rq2_main_improvements}
\renewcommand{\arraystretch}{1.15}
\noindent\makebox[\textwidth][l]{%
\resizebox{0.98\textwidth}{!}{%
\begin{tabular}{l lll lll}
\toprule
\multirow{2}{*}{\textbf{Method}} & \multicolumn{3}{c}{\textbf{Low-data setting}} & \multicolumn{3}{c}{\textbf{Full-data setting}} \\
\cmidrule(lr){2-4}\cmidrule(lr){5-7}
 & \textbf{Completion} & \textbf{Translation} & \textbf{Summarization} & \textbf{Completion} & \textbf{Translation} & \textbf{Summarization} \\
 & \textbf{Exact} & \textbf{Exact} & \textbf{METEOR} & \textbf{Exact} & \textbf{Exact} & \textbf{METEOR} \\
\midrule
Llama-3.2-1B Baseline & 75.42 & 38.28 & 25.83 & 81.94 & 56.94 & 28.91 \\
\rowcolor{gray!20} Llama-3.2-1B \TheName{} & 81.49 {\scriptsize(+8.0\%)} & 41.99 {\scriptsize(+9.7\%)} & 27.58 {\scriptsize(+6.8\%)} & 85.13 {\scriptsize(+3.9\%)} & 60.53 {\scriptsize(+6.3\%)} & 31.59 {\scriptsize(+9.3\%)} \\
\midrule
DeepSeek-Coder-1.3B Baseline & 67.91 & 44.50 & 30.23 & 74.43 & 66.87 & 30.79 \\
\rowcolor{gray!20} DeepSeek-Coder-1.3B \TheName{} & 78.60 {\scriptsize(+15.8\%)} & 49.28 {\scriptsize(+10.8\%)} & 30.65 {\scriptsize(+1.4\%)} & 99.32 {\scriptsize(+33.4\%)} & 69.98 {\scriptsize(+4.7\%)} & 32.40 {\scriptsize(+5.2\%)} \\
\midrule
StarCoderBase-1B Baseline & 73.67 & 37.44 & 30.53 & 74.13 & 58.25 & 31.75 \\
\rowcolor{gray!20} StarCoderBase-1B \TheName{} & 98.94 {\scriptsize(+34.3\%)} & 42.46 {\scriptsize(+13.4\%)} & 31.24 {\scriptsize(+2.3\%)} & 99.62 {\scriptsize(+34.4\%)} & 61.36 {\scriptsize(+5.3\%)} & 33.95 {\scriptsize(+6.9\%)} \\
\midrule
Qwen2.5-Coder-1.5B Baseline & 77.85 & 36.36 & 18.41 & 79.89 & 57.18 & 28.88 \\
\rowcolor{gray!20} Qwen2.5-Coder-1.5B \TheName{} & 96.81 {\scriptsize(+24.4\%)} & 42.94 {\scriptsize(+18.1\%)} & 20.71 {\scriptsize(+12.5\%)} & 99.70 {\scriptsize(+24.8\%)} & 62.08 {\scriptsize(+8.6\%)} & 30.86 {\scriptsize(+6.9\%)} \\
\midrule
SmolLM3-3B Baseline & 77.69 & 37.56 & 15.29 & 81.87 & 55.74 & 27.12 \\
\rowcolor{gray!20} SmolLM3-3B \TheName{} & 92.64 {\scriptsize(+19.2\%)} & 41.39 {\scriptsize(+10.2\%)} & 20.23 {\scriptsize(+32.3\%)} & 99.54 {\scriptsize(+21.6\%)} & 60.17 {\scriptsize(+7.9\%)} & 28.70 {\scriptsize(+5.8\%)} \\
\midrule
Granite-Code-2B Baseline & 71.93 & 37.68 & 23.81 & 73.44 & 55.50 & 32.20 \\
\rowcolor{gray!20} Granite-Code-2B \TheName{} & 74.66 {\scriptsize(+3.8\%)} & 40.67 {\scriptsize(+7.9\%)} & 27.39 {\scriptsize(+15.0\%)} & 77.31 {\scriptsize(+5.3\%)} & 61.48 {\scriptsize(+10.8\%)} & 33.07 {\scriptsize(+2.7\%)} \\
\bottomrule
\end{tabular}}}
\end{table*}

\paragraph{Structure-Preserving Code Tasks.}
The strongest evidence comes from completion and translation, where the target output preserves executable code structure. Completion gains are largest for code-specialized backbones after targeted recipe selection, reaching +34.4\% relative improvement while remaining positive for all other models. Translation is less peaked but more uniformly stable, with relative gains up to +18.1\% in the low-data setting and +10.8\% in the full-data setting. These patterns suggest that gaze-derived priors are especially useful when semantic attention can directly constrain valid code tokens.


\paragraph{Natural-Language Summarization.}
Summarization is more sensitive to decoding, tokenizer behavior, and LoRA recipe choice than the two code-structure tasks, because the target output no longer preserves the input program structure token by token. After excluding pathological outputs and applying the DeepSeek tokenizer fix, \TheName{} improves every summarization cell under METEOR, suggesting that gaze priors still provide useful semantic guidance even when code understanding must be expressed in natural language.

\paragraph{Backbone-Level View.}
Figure~\ref{fig:rq2_model_task_profile} shows why evaluating multiple backbones is more informative than reporting a larger table alone. The average relative gain is similar for StarCoderBase, Qwen2.5-Coder, and SmolLM3 (about 16\%), while Llama-3.2 and Granite-Code show smaller but still consistent gains (about 7--8\%). This spread supports a model-agnostic interpretation without hiding backbone sensitivity: code-specialized backbones benefit most on completion, whereas Granite and Llama indicate that the signal is not limited to the strongest completion-tuned models.

\paragraph{Model and Data-Regime Robustness.}
Across Table~\ref{tab:rq2_main_improvements}, the selected \TheName{} runs improve every task, backbone, and data regime under the matched metric. The results support the main claim under a conservative interpretation: human-attention priors provide consistent positive signal across architectures and data regimes, with the strongest effect on structure-preserving code generation and a smaller but positive effect on natural-language summarization. The breadth of this pattern is important because it rules out a single-model or single-task explanation. It also motivates the next two analyses, which ask which session source and which gaze-derived components account for the observed gains.

Taken together, RQ2 establishes the outcome-level effect. RQ3 and RQ4 then unpack this effect mechanistically: RQ3 varies the source of the human-attention prior, while RQ4 removes individual signal components from the full recipe.

\begin{figure}[t]
  \centering
  \includegraphics[width=0.92\columnwidth]{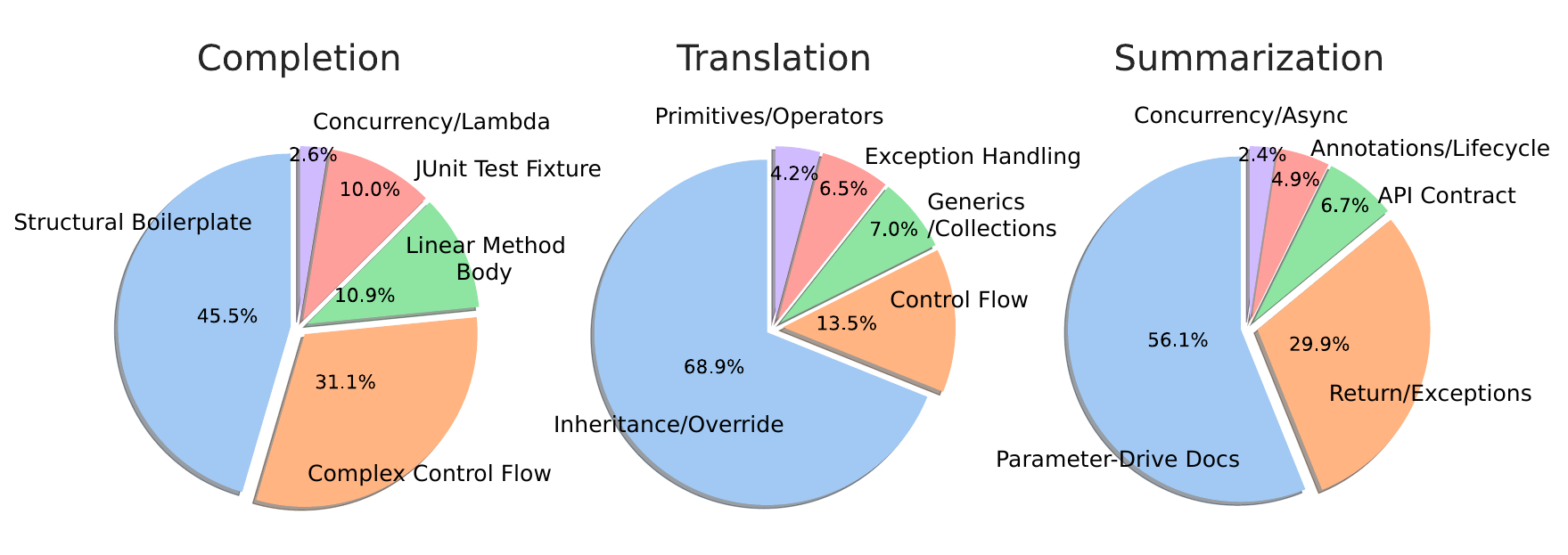}
  \caption{Semantic-category profiles across tasks. Completion emphasizes structural boilerplate, translation centers on inheritance/override patterns, and summarization shifts toward API and return semantics.}
  \label{fig:semantic_distribution}
\end{figure}

\begin{table}[t]
\caption{Session-mode results on current Llama-3.2-1B full-data runs. Gray cells mark the best variant per subcategory; parentheses show relative improvement over baseline.}
\label{tab:current_rq3_session_subcategories}
\centering
\scriptsize
\renewcommand{\arraystretch}{1.16}
\setlength{\tabcolsep}{1.8pt}
\resizebox{\columnwidth}{!}{%
\begin{tabular}{@{}l c c c c@{}}
\toprule
\textbf{Group} & \textbf{Base} & \textbf{Read} & \textbf{Write} & \textbf{Full} \\
\midrule
\multicolumn{5}{c}{\textbf{Completion} \quad (\textit{Exact})} \\
\midrule
Concurrency/Lambda & 64.46 & 67.47 {\scriptsize(+4.7\%)} & 70.48 {\scriptsize(+9.3\%)} & \cellcolor{gray!20}72.89 {\scriptsize(+13.1\%)} \\
Control Flow & 74.61 & 77.95 {\scriptsize(+4.5\%)} & \cellcolor{gray!20}79.51 {\scriptsize(+6.6\%)} & \cellcolor{gray!20}79.51 {\scriptsize(+6.6\%)} \\
Boilerplate & 98.89 & \cellcolor{gray!20}99.12 {\scriptsize(+0.2\%)} & \cellcolor{gray!20}99.12 {\scriptsize(+0.2\%)} & \cellcolor{gray!20}99.12 {\scriptsize(+0.2\%)} \\
Linear Body & 100.00 & 100.00 {\scriptsize(+0.0\%)} & 100.00 {\scriptsize(+0.0\%)} & 100.00 {\scriptsize(+0.0\%)} \\
JUnit Fixture & 72.60 & 74.43 {\scriptsize(+2.5\%)} & 74.43 {\scriptsize(+2.5\%)} & \cellcolor{gray!20}74.89 {\scriptsize(+3.1\%)} \\
\textbf{Overall} & 81.94 & 83.84 {\scriptsize(+2.3\%)} & 84.75 {\scriptsize(+3.4\%)} & \cellcolor{gray!20}85.13 {\scriptsize(+3.9\%)} \\
\midrule
\multicolumn{5}{c}{\textbf{Translation} \quad (\textit{Exact})} \\
\midrule
Multi-stmt Ctrl. & 13.93 & \cellcolor{gray!20}22.13 {\scriptsize(+58.8\%)} & 20.49 {\scriptsize(+47.1\%)} & 21.31 {\scriptsize(+52.9\%)} \\
Exceptions & 25.00 & 31.25 {\scriptsize(+25.0\%)} & 30.00 {\scriptsize(+20.0\%)} & \cellcolor{gray!20}33.75 {\scriptsize(+35.0\%)} \\
Generics/Coll. & 42.42 & 45.45 {\scriptsize(+7.1\%)} & \cellcolor{gray!20}48.48 {\scriptsize(+14.3\%)} & 45.45 {\scriptsize(+7.1\%)} \\
Inheritance & 25.00 & \cellcolor{gray!20}31.25 {\scriptsize(+25.0\%)} & \cellcolor{gray!20}31.25 {\scriptsize(+25.0\%)} & \cellcolor{gray!20}31.25 {\scriptsize(+25.0\%)} \\
Primitives/Ops & 72.99 & \cellcolor{gray!20}74.19 {\scriptsize(+1.6\%)} & 73.50 {\scriptsize(+0.7\%)} & 72.31 {\scriptsize(-0.9\%)} \\
\textbf{Overall} & 57.66 & \cellcolor{gray!20}60.53 {\scriptsize(+5.0\%)} & 59.81 {\scriptsize(+3.7\%)} & 59.33 {\scriptsize(+2.9\%)} \\
\midrule
\multicolumn{5}{c}{\textbf{Summarization} \quad (\textit{METEOR})} \\
\midrule
API Contract & 25.78 & 27.78 {\scriptsize(+7.8\%)} & \cellcolor{gray!20}28.01 {\scriptsize(+8.6\%)} & 27.81 {\scriptsize(+7.9\%)} \\
Async/Concur. & 25.85 & \cellcolor{gray!20}28.79 {\scriptsize(+11.3\%)} & 26.87 {\scriptsize(+3.9\%)} & 26.30 {\scriptsize(+1.7\%)} \\
Annotations & 29.72 & \cellcolor{gray!20}31.52 {\scriptsize(+6.1\%)} & 31.07 {\scriptsize(+4.6\%)} & 29.58 {\scriptsize(-0.5\%)} \\
Param Docs & 29.84 & 31.88 {\scriptsize(+6.8\%)} & 32.55 {\scriptsize(+9.1\%)} & \cellcolor{gray!20}32.61 {\scriptsize(+9.3\%)} \\
Return/Except. & 30.88 & \cellcolor{gray!20}34.53 {\scriptsize(+11.8\%)} & 34.17 {\scriptsize(+10.6\%)} & 34.38 {\scriptsize(+11.3\%)} \\
\textbf{Overall} & 28.91 & \cellcolor{gray!20}31.43 {\scriptsize(+8.7\%)} & 31.22 {\scriptsize(+8.0\%)} & 30.88 {\scriptsize(+6.8\%)} \\
\bottomrule
\end{tabular}%
}
\end{table}

\subsection{RQ3: Session-Mode Analysis}
\label{subsec:rq3}

Figure~\ref{fig:semantic_distribution} shows that semantic categories vary across tasks, motivating a comparison of reading-derived (\TheName{}(R)) and writing-derived (\TheName{}(W)) priors. We rerun this analysis on current Llama-3.2-1B full-data artifacts, using Exact Match for completion and translation and METEOR for summarization. Table~\ref{tab:current_rq3_session_subcategories} reports all baseline and prior variants by group.


\textbf{Completion subcategories.}
Writing-derived and full priors are strongest for code-production cases. The full variant leads on Concurrency/Lambda (+13.1\%), JUnit Test Fixture (+3.1\%), and completion (+3.9\%), while writing ties on Control Flow and Boilerplate. This suggests that writing traces capture generation-oriented dependencies, with the cleaned runs giving a more conservative estimate than earlier validation results.

\textbf{Translation and summarization subcategories.}
Reading-derived priors are strongest for translation overall (+5.0\%) and for Multi-statement Control Flow, Inheritance, and Primitives \& Operators. The full variant is strongest for Exceptions, and writing is strongest for Generics/Collections. Summarization is more mixed under METEOR: reading leads Async/Concurrency, Annotations, Return/Exception Semantics, and the overall aggregate, while writing leads API Contract and the full model leads Parameter Docs. These results indicate that comprehension-oriented reading priors help many input-understanding cases, but no single session mode dominates all natural-language summary categories.

\textbf{Combined session effect.}
The full-prior variant remains robust, especially for completion, but the current Llama results show a more nuanced pattern than the previous draft: specialized session priors often win individual rows. We therefore treat session mode as a task- and category-dependent design choice rather than as a universally monotonic improvement from combining reading and writing statistics.

\subsection{RQ4: Component Ablation}
\label{subsec:rq4}

RQ4 examines whether the full gaze signal matters beyond any single component. Table~\ref{tab:current_rq4_ablation} reports complete component-removal sweeps spanning all three tasks: completion on Llama-3.2-1B, and translation and summarization on DeepSeek-Coder and StarCoder. Removing either AST salience or rarity/transition flow reduces performance relative to the full recipe, and randomizing the weights also underperforms Full. This pattern supports the full \TheName{} recipe: semantic salience and sequential gaze flow are complementary rather than interchangeable.

\begin{table}[t]
\centering
\small
\setlength{\tabcolsep}{4pt}
\renewcommand{\arraystretch}{1.03}
\caption{Component ablation across complete variant sweeps. Completion and translation use Exact Match and summarization uses METEOR; drops are relative to the Full \TheName{} recipe.}
\label{tab:current_rq4_ablation}
\begin{tabular}{llrr}
\toprule
\textbf{Backbone / Task} & \textbf{Variant} & \textbf{Score} & \textbf{Drop} \\
\midrule
\multirow{4}{*}{\shortstack[l]{Llama\\Completion}} & Full & \cellcolor{gray!20}85.13 & -- \\
 & No semantic & 83.69 & -1.7\% \\
 & No rare/flow & 83.99 & -1.3\% \\
 & Random & 84.37 & -0.9\% \\
\midrule
\multirow{4}{*}{\shortstack[l]{DeepSeek\\Translation}} & Full & \cellcolor{gray!20}65.66 & -- \\
 & No semantic & 53.13 & -19.1\% \\
 & No rare/flow & 53.13 & -19.1\% \\
 & Random & 52.44 & -20.1\% \\
\midrule
\multirow{4}{*}{\shortstack[l]{DeepSeek\\Summ.}} & Full & \cellcolor{gray!20}33.92 & -- \\
 & No semantic & 30.11 & -11.2\% \\
 & No rare/flow & 30.22 & -10.9\% \\
 & Random & 30.22 & -10.9\% \\
\midrule
\multirow{4}{*}{\shortstack[l]{StarCoder\\Translation}} & Full & \cellcolor{gray!20}64.97 & -- \\
 & No semantic & 50.35 & -22.5\% \\
 & No rare/flow & 51.04 & -21.4\% \\
 & Random & 51.74 & -20.4\% \\
\midrule
\multirow{4}{*}{\shortstack[l]{StarCoder\\Summ.}} & Full & \cellcolor{gray!20}33.41 & -- \\
 & No semantic & 30.92 & -7.5\% \\
 & No rare/flow & 30.97 & -7.3\% \\
 & Random & 31.09 & -7.0\% \\
\bottomrule
\end{tabular}
\end{table}

\subsection{RQ5: Attention Distribution}
\label{subsec:rq5}

We treat attention distribution as diagnostic evidence rather than as part of the current six-backbone confirmation grid. Table~\ref{tab:current_attention_quality} recomputes the earlier attention-quality view on the selected current Llama-3.2-1B full-data adapters, using 24 held-out examples per task. The clearest current signal is improved generation confidence, especially for summarization; distributional attention measures are smaller and task-dependent, so we use them to interpret behavior rather than to replace the main quality results in Table~\ref{tab:rq2_main_improvements}.

\begin{table}[t]
\centering
\caption{Attention diagnostics on current Llama-3.2-1B full-data adapters. Arrows indicate the expected direction for each metric; gray cells mark the better score, and the table is used as diagnostic rather than primary quality evidence.}
\label{tab:current_attention_quality}
\resizebox{\columnwidth}{!}{%
\begin{tabular}{llccc}
\toprule
\textbf{Task} & \textbf{Metric} & \textbf{Baseline} & \textbf{\TheName{}} & \textbf{$p$-value} \\
\midrule
\multirow{4}{*}{Completion} & Generation Confidence $\uparrow$ & -0.0168 & \cellcolor{gray!20}-0.0123 & 3.78e-01 \\
 & Recency Focus $\uparrow$ & \cellcolor{gray!20}0.06942 & 0.06941 & 8.26e-04 \\
 & Avg. Focus $\uparrow$ & \cellcolor{gray!20}0.80347 & 0.80345 & 2.44e-01 \\
 & Entropy $\downarrow$ & 8.62056 & \cellcolor{gray!20}8.62055 & 2.18e-01 \\
\midrule
\multirow{4}{*}{Translation} & Generation Confidence $\uparrow$ & -0.0578 & \cellcolor{gray!20}-0.0407 & 1.80e-01 \\
 & Recency Focus $\uparrow$ & 0.35290 & \cellcolor{gray!20}0.35314 & 9.12e-03 \\
 & Avg. Focus $\uparrow$ & 0.76240 & \cellcolor{gray!20}0.76273 & 4.48e-03 \\
 & Entropy $\downarrow$ & \cellcolor{gray!20}5.94433 & 5.94471 & 5.81e-03 \\
\midrule
\multirow{4}{*}{Summarization} & Generation Confidence $\uparrow$ & -0.9126 & \cellcolor{gray!20}-0.6794 & 2.68e-03 \\
 & Recency Focus $\uparrow$ & \cellcolor{gray!20}0.18572 & 0.18563 & 2.80e-03 \\
 & Avg. Focus $\uparrow$ & \cellcolor{gray!20}0.73555 & 0.73555 & 9.16e-01 \\
 & Entropy $\downarrow$ & \cellcolor{gray!20}7.11796 & 7.11802 & 2.13e-01 \\
\bottomrule
\end{tabular}}
\end{table}

\begin{figure}[t]
     \centering
     \includegraphics[width=\columnwidth]{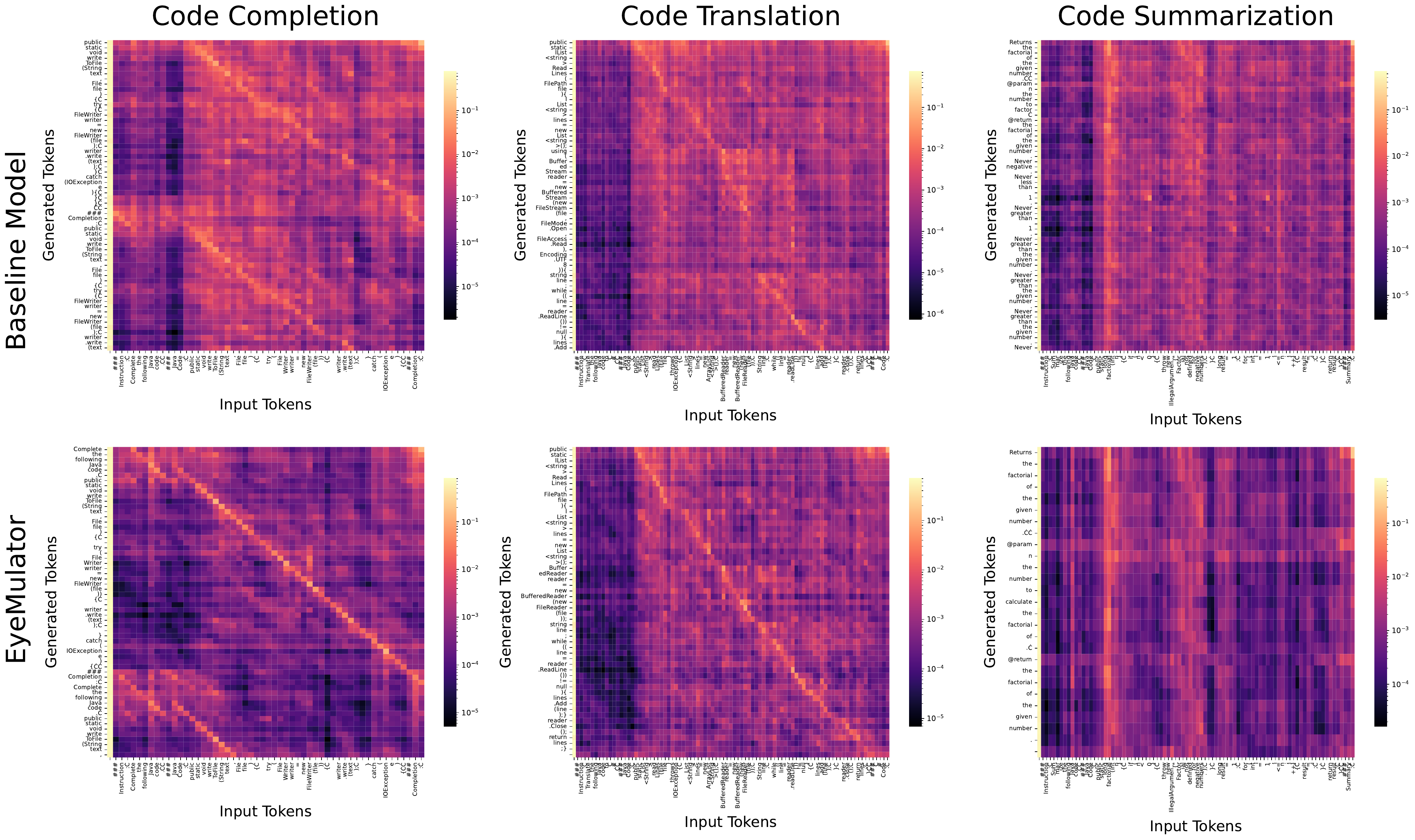}
     \caption{Qualitative attention diagnostic. Compared with the baseline (top), \TheName{} (bottom) concentrates more mass on semantically relevant input tokens, illustrating the intended effect of gaze-derived training priors.}
     \label{fig:master_attention_view}
\end{figure}

\paragraph{Qualitative Analysis.}
Figure~\ref{fig:master_attention_view} visualizes final-layer attention from generated tokens back to input tokens. We keep this qualitative view as an illustrative diagnostic rather than as final quantitative evidence. A broader selection of qualitative case studies is presented in Appendix~\ref{sec:appendix_examples}.

\section{Discussion}
\label{sec:discussion}

\TheName{} builds on a simple observation: developers do not process code uniformly. They revisit declarations, control-flow predicates, API contracts, and other semantic anchors while skimming boilerplate. Our results suggest that compact CodeLLMs benefit when this selectivity is distilled into training-time priors rather than architecture changes.


\paragraph{Why the gains are task-dependent.}
The largest improvements appear in completion and translation, where outputs preserve program structure and exact token choices matter. In these settings, gaze-derived weights help the model prioritize variables, signatures, and control-flow anchors that constrain valid code. Summarization is more diffuse because the output is natural language, making gains more dependent on recipe selection and tokenizer behavior. This helps explain why summarization gains are smaller yet still positive under METEOR.

\paragraph{Why modular priors matter.}
Because \TheName{} changes only the training objective and data weights, it can be applied across heterogeneous backbones without requiring a custom transformer. The component ablation further shows that semantic salience and transition/rarity flow are complementary: removing either signal consistently weakens the full recipe. This supports the view that human attention is useful not as a single heuristic, but as a structured prior combining what developers inspect and how they move through code.

\paragraph{Interpreting diagnostic evidence.}
The attention diagnostics serve as behavioral support, not the primary evaluation target. They show that gaze-informed fine-tuning can shift model behavior in interpretable ways, while downstream task quality remains the central evidence. Thus, \TheName{}’s main claim rests on improvements across the evaluation grid, with attention analyses explaining why those gains are plausible.

\section{Related Work}
\label{sec:related_work}

Prior work in code intelligence has advanced attention-based Transformers, yet often omits human cognitive cues about how developers actually inspect code. \TheName{} unifies these directions by integrating distilled attention priors, n-gram rarity weighting, and sequential gaze modeling into a single framework that can guide standard CodeLLM fine-tuning.

\subsection{Human-centered AI for Software Engineering}
Human-centered AI integrates cognitive insights to align models with developer workflows~\cite{abrahãoSoftwareEngineeringHumans2025,zhang2025enhancing,karas2024tale,li2024machines}. Eye-tracking has historically illuminated how programmers manage cognitive load~\cite{sharafi_practical_2020, sharafi_systematic_2015, grabinger_on_eye_2024, sharafi_eye-tracking_2015}, identifying key segments to improve automated summarization~\cite{bansal2023towards, Improving2014Rodeghero, sharafi_systematic_2015}. Recent work deepens this integration by correlating mouse interactions with neural attention~\cite{paltenghiThinkingDeveloperComparing2021}, training predictive gaze models~\cite{bansal2023modeling}, and incorporating gaze into transformer architectures~\cite{zhang2024eyetrans} or program repair~\cite{huberWhereLookWhen2023}. Unlike these approaches, \TheName{} distills gaze artifacts into modular, task-agnostic priors that can be injected into any pre-trained model without architectural changes, preserving sample efficiency.

\subsection{Large Language Models for Code Intelligence}
LLMs such as StarCoder~\cite{liStarCoderMaySource2023}, Llama-3.2~\cite{meta2024llama3_2, grattafioriLlama3Herd2024}, and DeepSeek-Coder~\cite{guoDeepSeekcoderWhenLarge2024} have advanced code generation~\cite{nam2024using,coignion2024performance,zhang2025codegrad,feng2020codebert}. Strategies to refine performance include Retrieval-Augmented Generation (RAG)~\cite{wang2025coderagbenchretrievalaugmentcode,yang2025empirical,guo2024retrieval,parvez2021retrieval}, instruction tuning~\cite{ouyangTrainingLanguageModels2022b}, and reasoning frameworks like SemCoder~\cite{ding2024semcodertrainingcodelanguage}. However, models still struggle with deep semantic understanding~\cite{nguyenEmpiricalStudyCapability2024a,he2024exploring,zhong2024can,yang2025elaboration}, leading to hallucinations~\cite{liuYourCodeGenerated2023a, siddiqQualityAssessmentChatGPT2024, zhangLLMHallucinationsPractical2025}. Existing feedback methods often lack token-level granularity~\cite{xuLLMRefinePinpointingRefining2024, douStepCoderImproveCode2024}. \TheName{} bridges this gap by injecting gaze-derived salience directly into self-attention, enhancing semantic grounding.

\subsection{Preference Learning and Model Alignment}
Preference learning aligns models with developer needs beyond simple correctness~\cite{jiangSurveyHumanPreference2024,slocum2025diverse,fang2025dpo,xiong2023iterative}. While Reinforcement Learning from Human Feedback (RLHF)~\cite{ouyangTrainingLanguageModels2022b, zhang2025leveraging,kirk2023understanding,wang2024comprehensive} is standard, direct optimization methods like DPO~\cite{rafailovDirectPreferenceOptimization2024a}, SimPO~\cite{mengSimPOSimplePreference2024}, and KTO~\cite{ethayarajhKTOModelAlignment2024} offer efficient alignment. Techniques such as Group Relative Policy Optimization (GRPO)~\cite{shao2024deepseekmathpushinglimitsmathematical} further stabilize training. \TheName{} extends this landscape by incorporating gaze-derived salience priors as token-level feedback within DPO, enabling precise alignment with human cognitive processes.

\section{Conclusion}
\label{sec:conclusion}

We present \TheName{}, a lightweight, model-agnostic framework that injects human gaze signals into CodeLLM fine-tuning. By mapping eye-tracking data from 27 programmers onto AST tokens, we derive semantic salience and gaze-transition priors, then incorporate them into parameter-efficient fine-tuning. Across six backbones, three tasks, and two data regimes, \TheName{} yields positive quality-safe gains in every evaluated cell, with the strongest effects on completion and translation and smaller but consistent gains on summarization. Session-mode and component-ablation analyses, together with appendix diagnostics, show that gaze-derived signals change model behavior in interpretable, task-dependent ways.

\section*{Acknowledgments}

We thank the study participants from Vanderbilt University and the University of Notre Dame for contributing to the human gaze data that supports this work. 
We also thank our collaborators and lab members for their feedback on the eye-tracking setup, artifact preparation, and manuscript presentation.

\section{Limitations}
\label{sec:limitations}

While \TheName{} demonstrates significant improvements in code intelligence tasks, we acknowledge several limitations in our current study, spanning model scale, language coverage, and participant demographics.

\paragraph{Model Scale and Compute.}
Due to computational resource constraints, our evaluation focuses on small code language models from roughly 1B to 3B parameters. This range allows broad coverage across six backbones and two data regimes, but does not establish scaling behavior for 7B, 13B, or larger systems.


\paragraph{Language and Task Diversity.}
Our priors are distilled from gaze recordings on \textbf{Java} code (EyeTrans) and evaluated on Java/C\# static code comprehension tasks, so transfer to structurally distinct languages (e.g., Python, Haskell) or markup (HTML/CSS), as well as to dynamic, interactive editing environments that require temporal gaze modeling, remains unverified.

\paragraph{Participant Demographics.}
The gaze patterns were distilled from a cohort of 27 verified programmers. While this sample size is consistent with prior eye-tracking research, it may not fully capture the cognitive diversity of the global developer population across different experience levels, neurodiverse traits, or cultural coding practices.

\section{Ethical Considerations}
\label{sec:ethics}

\paragraph{Potential for Misuse.}
Gaze analysis carries a dual-use risk in workplace surveillance. \TheName{} is designed only to distill \emph{aggregated} cognitive patterns for model training, not to assess or track individual developers, and we discourage any use that blurs this boundary.

\paragraph{Data Privacy and Dataset Usage.}
We use the publicly available \textit{EyeTrans} dataset under its original IRB-approved consent and anonymization protocols, and further mitigate re-identification risk by releasing only aggregated statistics (Beta priors and n-gram counts), not per-participant traces.

\paragraph{Use of AI Assistants.}
The authors used AI assistants (LLMs) for minor editorial assistance, including language polishing and \LaTeX{} formatting. All research ideation, experimental design, implementation, analysis, and claims are the authors' own.



\section*{Data Availability Statement}
\label{sec:data_availability}

The distilled human-attention artifacts, including gaze priors, a schema-matching dataset sample, and reference implementations, are archived on Zenodo at \url{https://zenodo.org/records/17205682}; the underlying eye-tracking data originates from the EyeTrans corpus~\cite{zhang2024eyetrans} and is redistributed under its original ethical-use terms.

\bibliography{custom}


\appendix

\section{Evaluation Metrics}
\label{sec:appendix_metrics}

To ensure reproducibility, we provide formal definitions for all metrics used to evaluate task performance (RQ2--RQ4).

\subsection{Task Performance Metrics}
\label{subsec:task_metrics}

\subsubsection{Code Translation and Completion}
For Java-to-C\# translation and code completion, the main tables report Exact Match, with additional code metrics available for audit:

\begin{itemize}
    \item \textbf{Exact Match:}
    A prediction receives credit when the whitespace-normalized prediction and reference are identical, or when one contains the other as a complete target span:
    \begin{equation}
        \text{Exact} = \mathbb{I}(\hat{y} = y \lor \hat{y} \in y \lor y \in \hat{y})
    \end{equation}
    where $y$ is the ground truth, $\hat{y}$ is the generated code, and $\mathbb{I}(\cdot)$ is the indicator function. This exact-style criterion is appropriate for code tasks where generations may include the target span plus surrounding syntactic context.

    \item \textbf{Hybrid Exact Match (H-Exact):} 
    To account for minor formatting variations while rewarding precision, we calculate a weighted average of strict exact match and substring inclusion:
    \begin{equation}
        \text{H-Exact} = 0.5 \times \mathbb{I}(y = \hat{y}) + 0.5 \times \mathbb{I}(\hat{y} \in y)
    \end{equation}
    
    \item \textbf{CodeBLEU:} 
    Unlike standard BLEU, CodeBLEU integrates syntactic and semantic properties. It is computed as the weighted sum of four components:
    \begin{equation}
    \begin{split}
        \text{CodeBLEU} &= w_1 \text{BLEU} + w_2 \text{BLEU}_{\text{weighted}} \\
        &\quad + w_3 \text{Match}_{\text{ast}} + w_4 \text{Match}_{\text{df}}
    \end{split}
    \end{equation}
    where $\text{BLEU}_{\text{weighted}}$ gives higher weight to keywords, $\text{Match}_{\text{ast}}$ measures Abstract Syntax Tree similarity, and $\text{Match}_{\text{df}}$ measures data-flow graph similarity.
    
    \item \textbf{CrystalBLEU:} 
    A variant of BLEU optimized for code that filters out ``trivially shared'' $n$-grams (e.g., frequent syntax like \texttt{public void}) to prevent inflated scores. It calculates $n$-gram precision only on a distinct set of $n$-grams not appearing in the top-$k$ most frequent occurrences in the training corpus.
\end{itemize}

\subsubsection{Code Summarization}
For Java-to-Natural Language summarization, the main tables report METEOR, with other standard text-generation metrics used for audit:

\begin{itemize}
    \item \textbf{METEOR:} Computes the harmonic mean of precision and recall, incorporating stemming and synonym matching to capture semantic overlap.
    \item \textbf{ROUGE-L:} Measures the Longest Common Subsequence (LCS) between the candidate summary and the reference, capturing sentence-level structure.
    \item \textbf{BERTScore:} Computes the similarity between candidate and reference summaries using contextual embeddings from a pre-trained BERT model:
    \begin{equation}
        R_{\text{BERT}} = \frac{1}{|x|} \sum_{x_i \in x} \max_{y_j \in y} \mathbf{x}_i^\top \mathbf{y}_j
    \end{equation}
    where $\mathbf{x}_i$ and $\mathbf{y}_j$ are the embedding vectors for tokens in the candidate and reference, respectively.
\end{itemize}

\section{Initial Validation Results}
\label{sec:appendix_initial_validation}

Table~\ref{tab:appendix_initial_validation_results} preserves the earlier result table. We include it as historical and diagnostic evidence that the method improved over matched baselines in the initial three-backbone study. The current main-text Table~\ref{tab:rq2_main_improvements} should be read as the final confirmation: it uses the cleaned six-backbone evaluation, two data regimes, current quality-safe recipe selection, and the final metric choices described in Appendix~\ref{sec:appendix_metrics}. The qualitative attention map in Figure~\ref{fig:master_attention_view} comes from the same initial diagnostic package and is retained to illustrate the attention-shift intuition.

\begin{table*}[t]
\centering
\small
\caption{Initial three-backbone validation results retained for context. The main-text Table~\ref{tab:rq2_main_improvements} reports the final six-backbone confirmation study used for the paper's main claims.}
\label{tab:appendix_initial_validation_results}
\renewcommand{\arraystretch}{1.15}
\resizebox{\textwidth}{!}{%
\begin{tabular}{l lll lll llll}
\toprule
\multirow{2}{*}{\textbf{Model}}
  & \multicolumn{3}{c}{\textbf{Completion}}
  & \multicolumn{3}{c}{\textbf{Translation}}
  & \multicolumn{4}{c}{\textbf{Summarization}} \\
\cmidrule(lr){2-4}\cmidrule(lr){5-7}\cmidrule(lr){8-11}
  & CodeBLEU & CrystalBLEU & H-Exact
  & CodeBLEU & CrystalBLEU & H-Exact
  & METEOR & ROUGE-1 & ROUGE-L & BERTScore \\
\midrule
StarCoder
  & 13.90 & 19.41 & 47.66
  & 52.07 & 65.07 & 21.11
  & 29.06 & 27.47 & 20.78 & 34.04 \\
\rowcolor{gray!20}
StarCoder\,(\TheName{})
  & 51.98\,{\scriptsize(+38.08)} & 57.53\,{\scriptsize(+38.12)} & 79.90\,{\scriptsize(+32.24)}
  & 86.42\,{\scriptsize(+34.35)} & 92.61\,{\scriptsize(+27.54)} & 64.97\,{\scriptsize(+43.86)}
  & 33.41\,{\scriptsize(+4.35)} & 46.27\,{\scriptsize(+18.80)} & 41.09\,{\scriptsize(+20.31)} & 51.06\,{\scriptsize(+17.02)} \\
\midrule
Llama-3.2
  & 19.45 & 26.65 & 50.85
  & 71.88 & 82.29 & 53.25
  & 27.05 & 24.86 & 18.27 & 33.41 \\
\rowcolor{gray!20}
Llama-3.2\,(\TheName{})
  & 60.47\,{\scriptsize(+41.02)} & 65.90\,{\scriptsize(+39.25)} & 77.96\,{\scriptsize(+27.11)}
  & 83.25\,{\scriptsize(+11.37)} & 91.52\,{\scriptsize(+9.23)} & 61.02\,{\scriptsize(+7.77)}
  & 30.69\,{\scriptsize(+3.64)} & 44.06\,{\scriptsize(+19.20)} & 38.84\,{\scriptsize(+20.57)} & 49.49\,{\scriptsize(+16.08)} \\
\midrule
DeepSeek-Coder
  & 13.80 & 18.98 & 49.40
  & 58.69 & 69.94 & 24.13
  & 22.81 & 21.96 & 15.12 & 28.64 \\
\rowcolor{gray!20}
DeepSeek-Coder\,(\TheName{})
  & 48.82\,{\scriptsize(+35.02)} & 54.63\,{\scriptsize(+35.65)} & 78.91\,{\scriptsize(+29.51)}
  & 86.34\,{\scriptsize(+27.65)} & 93.09\,{\scriptsize(+23.15)} & 65.66\,{\scriptsize(+41.53)}
  & 33.92\,{\scriptsize(+11.11)} & 46.86\,{\scriptsize(+24.90)} & 41.68\,{\scriptsize(+26.56)} & 51.56\,{\scriptsize(+22.92)} \\
\bottomrule
\end{tabular}%
}
\end{table*}

\section{Implementation Details}
\label{sec:appendix_implementation}

To ensure reproducibility, we provide the configuration for \TheName{} and the corresponding baselines across the six-backbone evaluation.

\subsection{Model and Loss Configuration}
We apply the same gaze-weighting mechanism to six backbones: \texttt{StarCoderBase-1B}~\cite{liStarCoderMaySource2023}, \texttt{Llama-3.2-1B}~\cite{meta2024llama3_2,grattafioriLlama3Herd2024}, \texttt{DeepSeek-Coder-1.3B}~\cite{guoDeepSeekcoderWhenLarge2024}, \texttt{Qwen2.5-Coder-1.5B}, \texttt{SmolLM3-3B}, and \texttt{Granite-Code-2B}. Each baseline uses standard causal language modeling, while \TheName{} scales per-token loss by gaze-derived weights $w_j$. The current experiments use LoRA/QLoRA adapters rather than full fine-tuning, which keeps the method lightweight and makes the six-model grid tractable.

\subsection{Hyperparameters}
We use task-level recipes rather than a single universal setting: completion and translation use rank-32 LoRA with weight 4, while summarization uses lower-rank variants selected from completed quality-safe sweeps. Table~\ref{tab:hyperparams} summarizes the main shared settings.

\begin{table}[h]
    \centering
    \caption{Main experimental hyperparameters for the current six-backbone LoRA/QLoRA evaluation.}
    \label{tab:hyperparams}
    \resizebox{\columnwidth}{!}{%
    \begin{tabular}{ll}
        \hline
        \textbf{Hyperparameter} & \textbf{Value} \\
        \hline
        Base Models & Six 1B--3B CodeLLM backbones \\
        Optimizer & AdamW \\
        LR Schedule & Linear with warmup \\
        Learning Rate & Recipe-dependent ($10^{-4}$ to $5\times10^{-5}$) \\
        Effective Batch Size & 16 \\
        Max Sequence Length & 1024 tokens \\
        Training Epochs & 3 \\
        Weight Decay & 0.01 \\
        DPO Weight $\gamma$ & 0.1 \\
        Base Token Weight $w_{\text{base}}$ & 1.0 \\
        Precision & bf16 / 4-bit QLoRA where supported \\
        Hardware & NVIDIA L40S-class GPUs \\
        \hline
    \end{tabular}
    }
\end{table}

\subsection{Attention Prior Processing}
Gaze-derived priors are injected into the loss function at the subword level. For each training example, we first run a language-specific AST parser over the source code and assign every leaf token a semantic label (e.g., \texttt{MethodDeclaration}, \texttt{IfStatement}, \texttt{VariableDeclarator}). Each label is then mapped to its posterior salience $\mathbb{E}[\theta_s]$ via the Beta priors distilled in Section~\ref{subsec:pattern_extraction}. When the tokenizer splits an AST token into multiple BPE subwords, the parent's salience is propagated to each shard. The final per-token weight $w_j$ combines this salience with a base term and an inverse-frequency correction (Section~\ref{subsec:gaze_informed_finetuning}). At training time, these weights are materialized into a single tensor that is broadcast against the cross-entropy loss, so that the procedure adds negligible overhead relative to standard fine-tuning.

For the \textit{Reading-derived} (\TheName{}(R)), \textit{Writing-derived} (\TheName{}(W)), and \textit{Full} variants in Table~\ref{tab:current_rq3_session_subcategories}, the only change is the source of the Beta and n-gram statistics (reading sessions, writing sessions, or both); the model, optimizer, and training schedule remain identical across all configurations.

\section{Qualitative Case Studies}
\label{sec:appendix_examples}

We present specific cases where \TheName{} demonstrates superior logic and conciseness compared to the baseline model.

\subsection{Case Study 1: Java-to-C\# Translation}
In this example, the baseline fails to terminate cleanly, producing a syntax error and unreachable code, while \TheName{} recognizes the high-level intent and uses a built-in C\# method.

\begin{figure}[!ht]
\centering
\begin{minipage}{\columnwidth}
\footnotesize
\noindent\textbf{Input (Java Code)}\vspace{-0.3em}
\begin{verbatim}
public int FindFirstRecordLocBySid(short sid) {
    int index = 0;
    for (IEnumerator iterator =
         records.GetEnumerator();
         iterator.MoveNext(); ) {
        Record record = (Record)iterator.Current;
        if (record.Sid == sid) { return index; }
        index++;
    }
    return -1;
}
\end{verbatim}
\vspace{-0.3em}
\noindent\textbf{Baseline (Llama-3.2-1B)}\vspace{-0.3em}
\begin{verbatim}
public int FindFirstRecordLocBySid(int sid) {
    int index = 0;
    foreach (Record record in records.Records) {
        if (record.Sid == sid) return index;
    }
    return -1;
} return -2;}
\end{verbatim}
\vspace{-0.3em}
\noindent\textbf{\TheName{} (Ours)}\vspace{-0.3em}
\begin{verbatim}
public int FindFirstRecordLocBySid(short sid) {
    int index = records.IndexOf(sid);
    return index;
}
\end{verbatim}
\end{minipage}
\caption{Translation case study. The baseline continues past a well-formed method and appends spurious code, while \TheName{} preserves the method intent and emits the idiomatic C\# \texttt{IndexOf} call with the correct \texttt{short} signature.}
\label{fig:case_study_translation}
\end{figure}

\subsection{Case Study 2: Code Summarization}
This case illustrates the ``repetition loop'' common in smaller LLMs: the baseline produces redundant Javadoc-style guarantees, whereas \TheName{} emits a single focused sentence.

\begin{figure}[!ht]
\centering
\begin{minipage}{\columnwidth}
\footnotesize
\noindent\textbf{Input (Java Method)}\vspace{-0.3em}
\begin{verbatim}
public String getSqlWithValues() {
    if( namedParameterValues.size() == 0 ) {
      return super.getSqlWithValues();
    }
    return result.toString();
}
\end{verbatim}
\vspace{-0.5em}
\noindent\textbf{Baseline Output.} \textit{Returns a SQL query with the named parameter values added. @return the SQL query. Never null. Always returns non empty. Never returns an empty string. Never returns a non-empty string\ldots\ [Truncated]}

\smallskip
\noindent\textbf{\TheName{} Output.} \textit{Returns the SQL query with the values of the parameters set.}
\end{minipage}
\caption{Summarization case study. The baseline falls into a repetitive Javadoc-style pattern, whereas \TheName{} produces a concise summary that captures the method's behavior.}
\label{fig:case_study_summarization}
\end{figure}



\section{Future Work}
\label{sec:appendix_future_work}

Several directions could extend \TheName{} beyond the current camera-ready scope. Future work should examine whether gaze-derived priors remain effective for stronger instruction-tuned and agentic models, transfer beyond Java-centered benchmarks to broader programming settings, capture richer temporal attention signals, and improve developer-facing coding workflows.

\textbf{Scaling to larger models.}
Our experiments focus on compact 1B--3B CodeLLMs, where full cross-task sweeps are tractable. Applying the same priors to larger instruction-tuned and agentic coding models would test whether gaze-derived signals still add value when base models already encode stronger code structure and reasoning patterns.

\textbf{Expanding language and task coverage.}
The present study uses Java-centered gaze artifacts and CodeXGlue-style tasks. Extending the pipeline to additional languages, multi-file contexts, and repository-level tasks would clarify whether the learned signal reflects Java-specific habits or broader program-structure priors.

\textbf{Modeling richer gaze dynamics.}
Our current priors summarize fixation and transition patterns, but eye-tracking data also contains dwell duration, regressions, scan-path uncertainty, and temporal shifts in attention. Incorporating these signals could support softer, context-dependent token weights that better distinguish quick skimming from deeper semantic inspection.

\textbf{Studying interactive use.}
Developer studies with \TheName{}-adapted models would complement offline metrics by evaluating practical effects in coding workflows. Such studies could measure whether gaze-informed outputs reduce review effort, debugging time, or cognitive load during real development tasks.

\end{document}